\documentclass[11pt]{article}
\pdfoutput=1 
\usepackage{amssymb}
\usepackage{amsmath}
\usepackage{amstext}
\usepackage{mathrsfs}
\usepackage{multirow}
\usepackage{graphicx}
\usepackage{epsfig}
\usepackage{verbatim} 
\usepackage{fancybox}
\usepackage{color}
\usepackage{mathdots}
\usepackage{anyfontsize}
\usepackage{ulem}
\usepackage{enumitem}
\usepackage{caption}
\usepackage{amsthm}
\usepackage{subfig}
\usepackage{mathtools}
\usepackage{tensor}
\usepackage{youngtab}
\usepackage{braket}
\usepackage{bbm}
\usepackage{parskip}
\usepackage[numbers,sort&compress]{natbib}
\usepackage{ytableau}
\usepackage[utf8]{inputenc}
\usepackage{tikz}
\usepackage{simpler-wick}
\usetikzlibrary{positioning, trees,decorations, decorations.markings, decorations.pathmorphing, arrows, shapes.geometric}

\newcommand{\Comment}[1]{{}}
\definecolor{darkblue}{rgb}{0.15,0.35,0.55}
\definecolor{reddish}{rgb}{0.65, 0.2, 0.2}
\usepackage[linktocpage=true]{hyperref}
\hypersetup{
colorlinks=true,
citecolor=darkblue,
linkcolor=reddish,
urlcolor=darkblue,
pdfauthor={},
pdftitle={},
pdfsubject={}
}

\usepackage{cleveref}

\numberwithin{equation}{section}

\makeatletter
\def\thickhline{%
  \noalign{\ifnum0=`}\fi\hrule \@height \thickarrayrulewidth \futurelet
   \reserved@a\@xthickhline}
\def\@xthickhline{\ifx\reserved@a\thickhline
               \vskip\doublerulesep
               \vskip-\thickarrayrulewidth
             \fi
      \ifnum0=`{\fi}}
\makeatother

\newlength{\thickarrayrulewidth}
\begin{document}

\renewcommand{\thefootnote}{\fnsymbol{footnote}}
~
\vspace{1.75truecm}
\thispagestyle{empty}
\begin{center}
{\LARGE \bf{
Fermionic Kaluza--Klein Reductions
}}
\end{center}

\vspace{1cm}
\centerline{\begingroup
\renewcommand{\thefootnote}{$\chi$}
\Large
	Francesco Bertucci\footnote{\href{mailto:fbertucc@olemiss.edu}{\texttt{fbertucc@olemiss.edu}}}
\endgroup
}
\vspace{.5cm}

\centerline{{\it Department of Physics and Astronomy,}}
 \centerline{{\it University of Mississippi, University, MS 38677}} 
 \vspace{.25cm}
 
\vspace{1cm}
\begin{abstract}
\noindent

We work out the Kaluza--Klein (KK) compactification on a general product spacetime for free spin-1/2 and spin-3/2 fields in arbitrary dimension. The analysis is performed at the level of the action, and the resulting lower-dimensional theory describes infinite towers of massive particles whose masses are related to the eigenvalues of the Dirac and Rarita--Schwinger operators on the internal space. A Hodge-like decomposition is used to write down a KK ansatz for spin-3/2 fields, and the fate of gauge invariance is discussed.

\end{abstract}

\newpage
\setcounter{tocdepth}{3}
\tableofcontents
\renewcommand*{\thefootnote}{\arabic{footnote}}
\setcounter{footnote}{0}

\newpage

\section{Introduction}

Precisely a century ago, in 1926, Klein wrote a seminal paper \cite{Klein:1926tv} to incorporate quantum mechanics in a proposal of Kaluza for the unification of gravity and electromagnetism \cite{Kaluza:1921tu}, in turn inspired by a work of Weyl \cite{Weyl:1918ib}. The most revolutionary feature of this set of ideas is the introduction of extra dimensions beyond the four that are commonly believed to govern our world. A compelling argument for the existence of extra dimensions is that they furnish an elegant mechanism to unify fundamental interactions, a long-standing problem in theoretical physics. Since then, the existence of physical theories defined on higher-dimensional spaces and their consequences have been widely studied, and have led to an impressive amount of results.

In the modern incarnation of Kaluza--Klein (KK) theory, a higher-dimensional theory is defined on the product of a lower-dimensional spacetime and a compact internal space. Extra dimensions are then integrated out, resulting in a lower-dimensional theory with an infinite tower of massive particles \cite{Salam:1981xd,Witten:1981me,Bailin:1987jd,Overduin:1997sri}. Remarkably, these massive states appear even when the higher-dimensional parent theory is massless. In this process of dimensional reduction, the masses in lower dimensions are related to the eigenvalues of certain differential operators defined on the internal space; if the parent theory is interacting, massive particles in lower dimensions will also interact, with coupling constants related to integrals of triple products of the corresponding eigenmodes. These differential operators are Laplacians for bosonic particles, whereas they are the Dirac operator or higher-spin generalizations for fermionic particles.

Despite the attractiveness of KK theory, there is in principle little indication of what the number of extra dimensions should actually be, and which geometry should accommodate them. The advent of supergravity \cite{Nath:1975nj,Freedman:1976xh,Freedman:1976py,Deser:1976eh,VanNieuwenhuizen:1981ae} suggested a limit of seven extra dimensions: that is because 11D is the maximum in which supersymmetry can exist, without having to introduce particles with spin higher than two in the supermultiplet \cite{Nahm:1977tg,Cremmer:1978km}. This explains the great interest towards the study of KK reductions of supergravity theories (see the reviews \cite{Duff:1986hr,Duff:2025tot}), where it is customary to truncate the spectrum to retain only the massless particles, or zero modes.

Supergravity was initially proposed as a way of combining quantum field theory and general relativity. According to supersymmetry, the graviton should be paired with a fermionic particle of spin 3/2, the gravitino.
The theory of a free spin-3/2 particle was described for the first time by Rarita and Schwinger \cite{Rarita:1941mf}. Arbitrary interactions with other fields are in general forbidden \cite{Velo:1969bt}, and their coupling to gravitons in higher dimensions can lead to gravitational anomalies \cite{Alvarez-Gaume:1983ihn}. Furthermore, a consistent theory which combines the presence of spin-2 and spin-3/2 fields must be supersymmetric \cite{Grisaru:1976vm}. Experimentally, massive spin-3/2 particles exist as resonances in QCD.

In this paper, we work out the complete KK reduction of massless free spin-1/2 and spin-3/2 fields on an arbitrary spacetime product $\mathcal{M}\times\mathcal{N}$, where $\mathcal{N}$ is a compact internal space which admits the existence of spinors. We write down a KK ansatz for the spinors on the internal space and work at the level of the action to derive the lower-dimensional theory. For spin-3/2 fields, we discuss the gauge symmetry of the action in higher dimensions, and its fate in the lower-dimensional theory. The presence of a cosmological constant and its consequences are also discussed. Some previous studies involving specific manifolds are \cite{Casher:1984ym,Fogleman:1984jd,Kim:1985ez}.

Our results --- together with their bosonic complement found in \cite{Hinterbichler:2013kwa} --- pave the way for the study of dimensional reduction of supergravity theories beyond the zero-mode truncation for general internal manifolds. In non-supersymmetric theories, recent studies have shown \cite{Bonifacio:2019ioc,Bertucci:2026janew} how violations of perturbative unitarity can be cured in KK theories by imposing certain sum rules which involve the eigenspectrum of the afore-mentioned differential operators on the internal space. More concretely, one considers the high-energy limit of tree-level $2\to2$ scattering amplitudes in the lower-dimensional theory, and this needs to agree with the behavior of the amplitude in higher dimensions. This matching is realized through the sum rules, which can thus be interpreted as geometrical constraints on the internal space $\mathcal{N}$.

These and similar sum rules were later derived purely from geometry, by using completeness of the eigenmode bases and expanding quadruple integrals of eigenmodes in different ways. The resulting system of sum rules is then amenable to a bootstrap analysis \cite{Belavin:1984vu,Rattazzi:2008pe,Poland:2018epd}, which is performed via semi-definite optimization \cite{Simmons-Duffin:2015qma}. The results are numerical bounds on the eigenvalues and triple integrals of eigenmodes for various geometries. These include closed Einstein and hyperbolic manifolds in different dimensions \cite{Bonifacio:2020xoc,Bonifacio:2021msa,Bonifacio:2021aqf,Radcliffe:2024jcg,Adve:2025rvf}. A representation theoretical approach was developed in \cite{Kravchuk:2021akc,Bonifacio:2023ban} for hyperbolic manifolds in two and three dimensions. Flat tori and orbifolds were considered in \cite{Bertucci:2026fjh}. Two-dimensional surfaces with spin were analyzed for the first time in this context in \cite{Gesteau:2023brw}.

This paper is organized as follows: in Section \ref{sec:2}, we perform the dimensional reduction of free massless spin-1/2 and spin-3/2 fields, and in the latter case we discuss gauge invariance on Ricci-flat and Einstein product spaces. We conclude in Section \ref{sec:concl}. The representation of Dirac matrices we use is discussed in Appendix \ref{app:clifford}, while in Appendix \ref{app:SpinDecomp} the mathematical details on the spin-3/2 decomposition are spelled out.

\paragraph{Notation and conventions.}

The signature of the metric is $(-,+,+,\dots,+)$.

Spacetime indices in $D=d+n$ dimensions are $M,N,\dots$ and run from 0 to $D-1$. We heavily make use of the decomposition $M=(\mu,p)$: indices in $d$ dimensions are $\mu,\nu,\dots$ and go from 0 to $d-1$. Indices in $n$ dimensions are $p,q,\dots$ and go from 1 to $n$.

Local Lorentz indices in $D$, $d$ and $n$ dimensions are $\mathscr{A},\mathscr{B},\dots$, $\mathfrak{a},\mathfrak{b},\dots$ and $\bar{\mathfrak{a}},\bar{\mathfrak{b}},\dots$, respectively.

Coordinates in $D$, $d$ and $n$ dimensions are $X$, $x$ and $y$, respectively. Integrals over the internal manifold are written as
$\int_\mathcal{N} \coloneq\int_\mathcal{N} d^ny\, \bar{e}$, with $\bar{e}^{\bar{\mathfrak{a}}}_p$ the curved vielbein on $\mathcal{N}$.

$I_D$ denotes the $2^{\lfloor D/2\rfloor}$-dimensional identity matrix in $D$ dimensions. Similarly for $d$ and $n$.

(Curved) Dirac matrices in $D$ and $d$ dimensions are denoted by $\Gamma^M$ and $\gamma^\mu$ respectively. Euclidean (curved) Dirac matrices in $n$ dimensions are denoted by $\tilde\Gamma^p$.

The various general relativity tensors or scalars in $D$ dimensions do not have any superscript, e.g. $R_{MN}$, $R$. In $d$ and $n$ dimensions they have a superscript, e.g. $R^{(d)}_{\mu\nu}$, $R^{(d)}$ and $R^{(n)}_{pq}$, $R^{(n)}$.

We (anti)symmetrize with weight 1.

\section{Kaluza--Klein analysis for free fermions}\label{sec:2}

Let $\mathcal{M}\times\mathcal{N}$ be a $D$-dimensional manifold ($D>2$) given by the product of a $d$-dimensional spacetime and an $n$-dimensional internal space $\mathcal{N}$. For later convenience, we further assume that $d$ is even\footnote{The odd-$d$ case can be analyzed with a different representation for the Dirac matrices in $D$ dimensions.} and $d>2$. $\mathcal{N}$ is a Riemannian compact spin manifold, which is also orientable and without boundary.

The full curved metric on $\mathcal{M}\times\mathcal{N}$ is given by
\begin{equation}
	ds^2 =G_{MN}(X)dX^MdX^N =g(x)_{\mu\nu}dx^\mu dx^\nu +\gamma(y)_{pq}dy^p dy^q,
\end{equation}
with $g_{\mu\nu}$ and $\gamma_{pq}$ the curved metrics on $\mathcal{M}$ and $\mathcal{N}$ respectively.

We work in the vielbein formalism. The $D$-dimensional vielbein $E^\mathscr{A}_M(X)$ is related to the flat metric $\eta_{\mathscr{AB}}$ by
\begin{equation}
	G_{MN} =E^\mathscr{A}_M E^\mathscr{B}_N\, \eta_{\mathscr{AB}}.
\end{equation}
The inverse vielbein is
\begin{equation}
	E^M_\mathscr{A} =G^{MN}\eta_{\mathscr{AB}}E^\mathscr{B}_N.
\end{equation}
Spacetime indices are raised and lowered with $G_{MN}$. Local Lorentz indices are raised and lowered with $\eta_{\mathscr{AB}}$. The determinant of the $D$-dimensional vielbein $E^\mathscr{A}_M$ is denoted by $E$. Dirac matrices with capital latin indices are functions of the spacetime, namely
\begin{equation}
	\Gamma^M(X) =E^M_\mathscr{A}(X)\Gamma^\mathscr{A},
\end{equation}
where $\Gamma^\mathscr{A}$ are numerical matrices (see Appendix \ref{app:clifford} for more details).

In this section, we derive the complete KK reduction for the $D$-dimensional theory of a free massless Dirac fermion, with spin 1/2 and 3/2.

\subsection{Spin 1/2}\label{sec:spin12}

As a warm-up, let us consider the theory of a free Dirac massless fermion $\lambda(X)$ on $\mathcal{M}\times\mathcal{N}$. The higher dimensional action is
\begin{equation}\label{eq:DiracDdim}
	S =-\int d^DX\,E\,i\,\bar\lambda\,\Gamma^M\nabla_M \lambda.
\end{equation}

To perform the dimensional reduction down to $d$ dimensions, we split $\mathscr{A}=(\mathfrak{a},\bar{\mathfrak{a}})$ and use the decomposition of $D$-dimensional Dirac matrices $\Gamma^\mathscr{A}$ (see Appendix \ref{app:clifford})
\begin{equation}\label{eq:gammamatrixtext}
	\begin{split}
		\Gamma^\mathfrak{a} &=\gamma^\mathfrak{a} \otimes I_n, \qquad \mathfrak{a}=0,\dots,d-1, \\
		\Gamma^{\bar{\mathfrak{a}}} &=\gamma_d \otimes \tilde\Gamma^{\bar{\mathfrak{a}}}, \qquad \bar{\mathfrak{a}}=1,\dots,n.
	\end{split}
\end{equation}
Here, $\gamma^\mathfrak{a}$ are the Dirac matrices in $d$ dimensions and they satisfy the usual (Lorentzian) Clifford algebra, while $\tilde\Gamma^{\bar{\mathfrak{a}}}$ are Dirac matrices in $n$ dimensions and satisfy the Euclidean Clifford algebra. $\gamma_d$ is the chirality matrix in $d$ dimensions.

The $D$-dimensional spinor $\lambda(X)$ can be decomposed as
\begin{equation}\label{eq:KKspinor}
	\lambda =\sum_i \psi_i(x) \otimes \chi_i(y),
\end{equation}
where $\psi_i(x)$ are spinors in $d$ dimensions and $\chi_i(y)$ form a basis of orthonormal eigenspinors of the Dirac operator on $\mathcal{N}$, namely
\begin{equation}\label{eq:spinorBasis}
	i\tilde\Gamma^{p}\nabla_p \chi_i =\lambda_i\chi_i, \quad\quad \int_\mathcal{N} \chi_i^\dagger\chi_j =\delta_{ij},
\end{equation}
with $\lambda_i\in\mathbb{R}$.

In writing \eqref{eq:KKspinor}, we include the zero modes --- or harmonic spinors --- with $\lambda_i=0$. Throughout this paper and when needed, we use the subscripts $\alpha,\beta,\dots$ to label harmonic spinors, i.e. $\chi_\alpha$.

By making use of the KK decomposition \eqref{eq:KKspinor} and integrating over the internal manifold $\mathcal{N}$ with the help of \eqref{eq:spinorBasis}, we get the $d$-dimensional action
\begin{equation}\label{eq:4dDiracPre}
	S =-\int d^dx\,e\sum_i \bar\psi_i \big(i\gamma^\mu \nabla_\mu +\lambda_i\gamma_d\big)\psi_i,
\end{equation}
where $e$ denotes the determinant of the $d$-dimensional vielbein $e^\mathfrak{a}_\mu(x)$ and $\gamma^\mu(x)=e_\mathfrak{a}^\mu(x)\gamma^\mathfrak{a}$. With the aid of the chirality transformation
\begin{equation}
	\psi_i \to e^{i\frac{3\pi}4\gamma_d}\psi_i =-\frac{1}{\sqrt{2}}\big( I_d -i\gamma_d \big)\psi_i
\end{equation}
we can get the more familiar form:
\begin{equation}\label{eq:4dFinalDirac}
	S =-\int d^dx\,e\sum_i \bar\psi_i \big(i\gamma^\mu \nabla_\mu -i\lambda_i\big)\psi_i.
\end{equation}

The spectrum of the theory is manifest, and the field content is:
\begin{itemize}
	\item one massless spin-1/2 Dirac field for each harmonic internal spinor;
	\item one massive spin-1/2 Dirac field for each non-harmonic eigenmode of the Dirac operator, with mass $m_i=|\lambda_i|$.
\end{itemize}
There are neither tachyons nor ghosts.

Let us comment on the existence of harmonic spinors on the internal space $\mathcal{N}$. There is a crucial difference compared to the bosonic case of \cite{Hinterbichler:2013kwa}: a spin manifold whose scalar curvature is $R^{(n)}\ge0$ and strictly positive somewhere does not admit harmonic spinors \cite{Hitchin:1974rbi}. This is essentially due to the Lichnerowicz--Weitzenb\"ock formula \eqref{eq:LWformula}. Therefore, the massless zero modes are not necessarily present in the $d$-dimensional theory. Moreover, if $R^{(n)}$ is identically zero, every harmonic spinor $\chi_\alpha$ is parallel, i.e. it satisfies
\begin{equation}
	\nabla_p\chi_\alpha=0.
\end{equation}
Parallel spinors are relatively rare. If a spin manifold admits a parallel spinor, then it is Ricci-flat. The existence of parallel spinors can also be related to the holonomy group of the spin manifold. To conclude, the Dirac spectrum on the spin manifold $\mathcal{N}$ --- and thus in particular the harmonic sector --- depends on the choice of spin structure on $\mathcal{N}$ \cite{baer2000}. There are as many possible spin structures as elements of the first cohomology group $H^1(\mathcal{N}, \mathbb{Z}_2)$.

The analysis of harmonic spinors of the Dirac operator has a long and rich history in the KK physics literature, due to its connections to the search of a unified theory and to supersymmetry breaking: see for instance \cite{Witten:1981me,Mecklenburg:1981ju,Candelas:1985en}.

\subsection{Spin 3/2 on Ricci-flat manifolds}\label{sec:spin32}

Let us now turn our attention to spin-3/2 fields. After writing down a KK ansatz for the internal components of a $D$-dimensional spin-3/2 Dirac field, we discuss gauge symmetry and derive the $d$-dimensional theory.

\subsubsection{Dimensional reduction}

Consider now a free massless Dirac spin-3/2 field $\Psi_M(X)$ on a Ricci-flat product manifold $\mathcal{M}\times\mathcal{N}$, i.e.
\begin{equation}\label{eq:RicciFlatD}
	R_{MN}=0.
\end{equation}
Its higher dimensional action is the Rarita-Schwinger (RS) action \cite{Rarita:1941mf}
\begin{equation}\label{eq:RS10daction}
	S_{RS} =-\int d^DX\,E\,i\,\bar\Psi_M \Gamma^{MNP}\nabla_N\Psi_P.
\end{equation}

The action \eqref{eq:RS10daction} is invariant under the gauge transformations\footnote{We work under the assumption that the spin connection is free from torsion.}
\begin{equation}\label{eq:RSgauge10d}
	\delta\Psi_M =\nabla_M \mathcal{E},
\end{equation}
where the $D$-dimensional gauge parameter $\mathcal{E}(X)$ is a Dirac spinor. Note that \eqref{eq:RicciFlatD} implies
\begin{align}
	R^{(d)}_{\mu\nu} &=0, \\
	R^{(n)}_{pq} &=0, \label{eq:NRicciFlat}
\end{align}
namely $\mathcal{M}$ and $\mathcal{N}$ are also Ricci-flat.

As in the spin-1/2 case, we need the appropriate KK decomposition for $\Psi_M(X)$ on a basis of operators on the internal space $\mathcal{N}$. Such a decomposition is available when $\mathcal{N}$ is Einstein, which is a weaker condition than \eqref{eq:NRicciFlat}. Let us take $n>2$, and we will comment on $n=1,2$ later. We can thus proceed to write \cite{Wang_1991,Homma_2019} (see Appendix \ref{app:SpinDecomp} for details)
\begin{equation}\label{eq:HodgeSpin32}
	\begin{split}
		\Psi_M &=\begin{pmatrix}
			\Psi_\mu(x,y) \\ \Psi_p(x,y)
		\end{pmatrix} \\
		&\quad =\begin{pmatrix}
			\sum_i \psi_{\mu,i}(x) \otimes \chi_i(y) \\
			\sum_\Omega \Phi_\Omega(x) \otimes \Psi_{p,\Omega}(y) +\sum_i \Psi_i(x) \otimes \left( \nabla_p -\frac1n \tilde\Gamma_p \tilde\Gamma^q \nabla_q \right)\chi_i(y) +\sum_i \tilde\Psi_i(x) \otimes \frac in \tilde\Gamma_p\chi_i(y)
		\end{pmatrix} \\
		&\quad\quad =\begin{pmatrix}
			\sum_i \psi_{\mu,i}(x) \otimes \chi_i(y) \\
			\sum_\Omega \Phi_\Omega(x) \otimes \Psi_{p,\Omega}(y) +\sum_i \Psi_i(x) \otimes \nabla_p \chi_i(y) +\sum_i \Omega_i(x) \otimes \frac in \tilde\Gamma_p\chi_i(y)
		\end{pmatrix},
	\end{split}
\end{equation}
where we have defined
\begin{equation}
	\Omega_i \coloneq\tilde\Psi_i +\lambda_i \Psi_i
\end{equation}
and $\chi_i$ is the spin-1/2 eigenbasis on $\mathcal{N}$ satisfying \eqref{eq:spinorBasis}. The coefficients of the expansions \eqref{eq:HodgeSpin32} are $d$-dimensional fields: $\psi_{\mu,i}$ are Dirac spin-3/2 fields, while $\Phi_\Omega$, $\Psi_i$ and $\tilde\Psi_i$ (and thus $\Omega_i$) are Dirac spin-1/2 fields.

The basis of spinor-valued 1-forms $\Psi_{p,\Omega}$ in $n$ dimensions satisfies the RS eigenvalue problem
\begin{equation}\label{eq:PsipOmegaEqs}
	i\tilde\Gamma^q\nabla_q \Psi_{p,\Omega} =\lambda_\Omega \Psi_{p,\Omega}, \quad\quad	\tilde\Gamma^p \Psi_{p,\Omega} =\nabla^p \Psi_{p,\Omega} =0, \quad\quad \int_\mathcal{N} \Psi^\dagger_{p,\Omega}\Psi^p_{\Omega'} =\delta_{\Omega\Omega'},
\end{equation}
with $\lambda_\Omega\in\mathbb{R}$.

The sums in \eqref{eq:HodgeSpin32} include zero modes with $\lambda_\Omega=0$ and $\lambda_i=0$; when needed, we will use the subscripts $\omega$ and $\alpha$ to label the zero modes $\Psi_{p,\omega}$ and $\chi_\alpha$, respectively. Note that since $\mathcal{N}$ is Ricci-flat, harmonic spinors are also parallel --- i.e. $\nabla_p\chi_\alpha=0$ --- and therefore the $\Psi_i(x)$ are only defined for $i\ne\alpha$, or equivalently the $\Psi_i(x)$ sum in \eqref{eq:HodgeSpin32} can be restricted to non-harmonic spinors only. Importantly, this will not be true in Section \ref{sec:spin32Ein} below.

For a recent study on the RS operator spectrum, see \cite{Homma_2019}. In particular, even if Lichnerowicz--Weitzenb\"ock-like formulas exist, they are not powerful enough to inform us on the existence of harmonic spinors.

\subsubsection{Gauge fixing}

Let us derive the gauge transformations in $d$ dimensions. The $D$-dimensional spinor parameter $\mathcal{E}(X)$ in \eqref{eq:RSgauge10d} is decomposed as
\begin{equation}\label{eq:RSgaugePar}
	\mathcal{E} =\sum_i \epsilon_i(x) \otimes \chi_i(y),
\end{equation}
where $\epsilon_i(x)$ are $d$-dimensional spinor gauge parameters. The gauge transformations in $d$ dimensions are
\begin{equation}
	\begin{split}
		\delta\psi_{\mu,i} &=\nabla_\mu\epsilon_i, \\
		\delta\Phi_\Omega &=0, \\
		\delta\Psi_i &=\epsilon_i, \\
		\delta\Omega_i &=0.
	\end{split}
\end{equation}
Therefore the fields $\Psi_i$ (which only exist for $\lambda_i\ne0$) are pure St\"uckelberg, while $\Omega_i$ are gauge invariant.

We can fix the gauge symmetry by choosing the unitary gauge
\begin{equation}
	\Psi_i =0.
\end{equation}
Therefore, for $\lambda_i=0$ the transformation
\begin{equation}
	\delta\psi_{\mu,\alpha} =\nabla_\mu\epsilon_\alpha
\end{equation}
survives as a $d$-dimensional gauge symmetry, and we will have a massless spin-3/2 field for each harmonic spinor.
In this gauge, the resulting $d$-dimensional action reads
\begin{equation}\label{eq:SRpreShift}
	\begin{split}
		S_{RS} &=\int d^dx\,e \bigg[ \sum_\Omega \Big(
		-\bar\Phi_\Omega i\gamma^\nu\nabla_\nu \Phi_\Omega
		-\lambda_\Omega\bar\Phi_\Omega\gamma_d\Phi_\Omega \Big) \\
		&\quad\quad\quad\quad\quad\quad +\sum_i \bigg( -i\,\bar\psi_{\mu,i}\gamma^{\mu\nu\rho} \nabla_\nu \psi_{\rho,i} +\lambda_i\,\bar\psi_{\mu,i}\gamma_d\gamma^{\mu\nu}\psi_{\nu,i} \\
		&\quad\quad\quad\quad\quad\quad\quad  +\frac{n-1}{n}\Big(\bar\Omega_i i\gamma^\mu \nabla_\mu \Omega_i -\lambda_i\frac{n-2}{n}\bar\Omega_i\gamma_d\Omega_i \Big) \\
		&\quad\quad\quad\quad\quad\quad\quad\quad +\bar\psi_{\mu,i}\,\gamma_d\gamma^{\mu\nu}\nabla_\nu \Omega_i
		-\bar\Omega_i\,\gamma_d\gamma^{\mu\nu}\nabla_\mu \psi_{\nu,i} \\
		&\quad\quad\quad\quad\quad\quad\quad\quad\quad -i\lambda_i\frac{n-1}{n}\big(\bar\psi_{\mu,i}\gamma^\mu \Omega_i -\bar\Omega_i\gamma^\mu \psi_{\mu,i}  \big) \bigg)\bigg].
	\end{split}
\end{equation}

At this stage, the kinetic term for $\Omega_i$ has the ``wrong'' sign: this is an artifact of the choice of gauge, and will be fixed in the final (diagonalized) form of the action. To do so, let us perform the following shifts on the massless and massive terms:
\begin{equation}\label{eq:RSshift}
	\begin{split}
		\psi_{\mu,\alpha} &\,\,\to\,\, \psi_{\mu,\alpha} -\frac{i}{d-2}\gamma_\mu\gamma_d\,\Omega_\alpha, \\
		\psi_{\mu,i} &\,\,\to\,\, \psi_{\mu,i} -\bigg( \frac{i(n-1)}{n(d-1)}\gamma_\mu\gamma_d +\frac{D-2}{n(d-1)\lambda_i}\nabla_\mu \bigg)\Omega_i, \quad \lambda_i\ne0,
	\end{split}
\end{equation}
followed by the chirality rotations
\begin{equation}
	\Phi_\Omega \,\,\to\,\, e^{i\frac{3\pi}4\gamma_d}\Phi_\Omega, \quad\quad \psi_{\mu,i} \,\,\to\,\, e^{i\frac\pi4\gamma_d}\psi_{\mu,i}, \quad\quad \Omega_i \,\,\to\,\, e^{i\frac\pi4\gamma_d}\Omega_i,
\end{equation}
where $e^{i\theta\gamma_d} =\cos\theta\, I_d+i\sin\theta\,\gamma_d$. The action thus becomes
\begin{equation}
	\begin{split}
		S_{RS} &=-\int d^dx\,e \bigg[ \sum_\Omega \Big(
		\bar\Phi_\Omega i\gamma^\mu\nabla_\mu \Phi_\Omega
		-i\lambda_\Omega\,\bar\Phi_\Omega\Phi_\Omega \Big) \\
		&\quad\quad\quad\quad\quad\quad +\sum_\alpha \bigg( i\,\bar\psi_{\mu,\alpha}\gamma^{\mu\nu\rho} \nabla_\nu \psi_{\rho,\alpha} +\frac{D-2}{n(d-2)}\bar\Omega_\alpha i\gamma^\mu\nabla_\mu\Omega_\alpha \bigg) \\
		&\quad\quad\quad\quad\quad\quad\quad +\sum_{i\ne\alpha} \bigg( i\,\bar\psi_{\mu,i}\gamma^{\mu\nu\rho} \nabla_\nu \psi_{\rho,i} -i\lambda_i\,\bar\psi_{\mu,i}\gamma^{\mu\nu}\psi_{\nu,i} \\
		&\quad\quad\quad\quad\quad\quad\quad\quad\quad\quad\quad +\frac{(n-1)(D-2)}{n^2(d-1)}\Big(\bar\Omega_i i\gamma^\mu\nabla_\mu\Omega_i -i\lambda_i\,\bar\Omega_i\Omega_i \Big) \bigg) \bigg],
	\end{split}
\end{equation}
where the $\Phi_\Omega$ sum includes zero modes but we have separated the zero modes and the massive modes for the Dirac eigenmodes on the internal space $\mathcal{N}$.

To conclude, we just need the renormalizations (recall that we are considering $n>2$)
\begin{equation}
	\begin{split}
		\Omega_\alpha &\,\,\to\,\, \sqrt{\frac{n(d-2)}{D-2}}\Omega_\alpha, \\
		\Omega_i &\,\,\to\,\, \sqrt{\frac{n^2(d-1)}{(n-1)(D-2)}}\Omega_i, \quad \lambda_i\ne0.
	\end{split}
\end{equation}
The final form of the action is (we unified the massless and massive sums again)
\begin{equation}\label{eq:RSFinalKK}
	\begin{split}
		S_{RS} &=-\int d^dx\,e \bigg[ \sum_\Omega
		\bar\Phi_\Omega \big(i\gamma^\mu\nabla_\mu
		-i\lambda_\Omega\big)\Phi_\Omega \\
		&\quad\quad\quad\quad\quad\quad +\sum_{i} \bigg( \bar\psi_{\mu,i}\big( i\gamma^{\mu\nu\rho} \nabla_\nu -i\lambda_i\gamma^{\mu\rho}\big)\psi_{\rho,i} +\bar\Omega_i\big( i\gamma^\mu\nabla_\mu -i\lambda_i\big)\Omega_i \bigg) \bigg].
	\end{split}
\end{equation}

The spectrum of the theory is now manifest, with
\begin{itemize}
	\item one massless spin-3/2 Dirac field $\psi_{\mu,\alpha}$ for each harmonic internal spinor;
	\item one massive spin-3/2 Dirac field $\psi_{\mu,i}$ for each non-harmonic eigenmode of the Dirac operator, with mass $m_i=|\lambda_i|$;
	\item one massless spin-1/2 Dirac field $\Omega_\alpha$ for each harmonic internal spinor;
	\item one massive spin-1/2 Dirac field $\Omega_i$ for each non-harmonic eigenmode of the Dirac operator, with mass $m_i=|\lambda_i|$;
	\item one massless spin-1/2 Dirac field $\Phi_\omega$ for each harmonic internal spinor-valued 1-form;
	\item one massive spin-1/2 Dirac field $\Phi_\Omega$ for each non-harmonic eigenmode of the RS operator, with mass $m_\Omega=|\lambda_\Omega|$;
\end{itemize}
There are neither tachyons nor ghosts. The final $d$-dimensional action has the expected residual gauge symmetry for massless spin-3/2 fields, which are in correspondence with harmonic spinors in $n$ dimensions.

As promised, let us comment on the $n=2$ and $n=1$ cases (see also Appendix \ref{app:SpinDecomp}).
\paragraph{Special case: $\boldsymbol{n=2}$.}
When $n=2$, the $\Phi_\Omega(x)$ sum in \eqref{eq:HodgeSpin32} is restricted to harmonic spinor-valued 1-forms $\Phi_\omega(x)$ only. This is the only difference, all the formulas are valid for any $n\ge2$ and the final result \eqref{eq:RSFinalKK} is valid, with $\lambda_\Omega=0$. Note in particular that the mass term for $\Omega_i(x)$ in \eqref{eq:SRpreShift} is zero for $n=2$, but a mass term is nevertheless generated through the shifts \eqref{eq:RSshift}.

\paragraph{Special case: $\boldsymbol{n=1}$.} When $n=1$, the divergence-free gamma traceless sector is absent, as well as the ``twistor'' sector, and we are left with
\begin{equation}
	\Psi_M =\begin{pmatrix}
		\Psi_\mu(x,y) \\ \Psi_{p=1}(x,y)
	\end{pmatrix} =\begin{pmatrix}
		\sum_i \psi_{\mu,i}(x) \otimes \chi_i(y) \\
		\sum_i \tilde\Psi_i(x) \otimes i\,\tilde\Gamma_1\chi_i(y),
	\end{pmatrix}
\end{equation}
for $\tilde\Gamma_1=\pm1$. The $d$-dimensional gauge transformations read
\begin{equation}
	\begin{split}
		\delta\psi_{\mu,i} &=\nabla_\mu\epsilon_i, \\
		\delta\tilde\Psi_i &=-\lambda_i\epsilon_i.
	\end{split}
\end{equation}
This time, $\tilde\Psi_i$ is a St\"uckelberg field for $\lambda_i\ne0$, and the unitary gauge is
\begin{equation}
	\tilde\Psi_i =0, \quad\quad \lambda_i\ne0.
\end{equation}
Once again, for $\lambda_i=0$ we have the surviving gauge symmetry
\begin{equation}
	\delta\psi_{\mu,\alpha} =\nabla_\mu\epsilon_\alpha.
\end{equation}

After algebraic manipulations (including a shift in the massless sector) the action for $n=1$ becomes
\begin{equation}
	S_{RS} =-\int d^dx\,e \bigg[ \sum_\alpha \bigg( i\,\bar\psi_{\mu,\alpha}\gamma^{\mu\nu\rho} \nabla_\nu \psi_{\rho,\alpha} +\bar{\tilde\Psi}_\alpha i\gamma^\mu\nabla_\mu\tilde\Psi_\alpha \bigg) +\sum_{i\ne\alpha} \bigg( \bar\psi_{\mu,i}\big(i\,\gamma^{\mu\nu\rho} \nabla_\nu -i\lambda_i\gamma^{\mu\rho}\big)\psi_{\rho,i} \bigg) \bigg].
\end{equation}

\subsection{Spin 3/2 on Einstein spaces}\label{sec:spin32Ein}

The massless spin 3/2 analysis assumed Ricci-flatness of the higher dimensional space $\mathcal{M}\times\mathcal{N}$. Let us now generalize this to massless spin-3/2 fields defined on $D$-dimensional Einstein spaces with negative cosmological constant \cite{Deser:1977uq,VanNieuwenhuizen:1981ae,Weinberg:2000cr,Buchbinder:2010gp,Rahman:2011ik}.

The $D$-dimensional action for a massless spin-3/2 field on a $D$-dimensional Einstein spin manifold with negative cosmological constant is
\begin{equation}\label{eq:mRSaction}
	S_{mRS} =-\int d^{D}X\,E\,\bar\Psi_M\bigg(i\,\Gamma^{MNP}\nabla_N -i\frac{D-2}{2L} \Gamma^{MP} \bigg)\Psi_P \eqcolon S_{RS} +S^{(m)}_{RS},
\end{equation}
where $L$ has the dimension of a length and it is defined through \eqref{eq:cosmcost} below.

The metric on the space $\mathcal{M}\times\mathcal{N}$ satisfies the Einstein equations
\begin{equation}
	R_{MN}-\frac12 R\,G_{MN} +\Lambda \, G_{MN}=0,
\end{equation}
with a negative cosmological constant\footnote{To include $D$-dimensional Einstein spaces with positive cosmological constant, we would either need to analytic continue $L\to iL$, which however spoils the hermiticity of the action \eqref{eq:mRSaction} \cite{Deser:2001pe,Deser:2001us}, or to give up gauge invariance, see e.g. \cite{Bergshoeff:2015tra}.}
\begin{equation}\label{eq:cosmcost}
	\Lambda =-\frac{(D-2)(D-1)}{2L^2}.
\end{equation}
In this case the action is invariant under the $D$-dimensional gauge transformations
\begin{equation}\label{eq:spin32gauge10d}
	\delta\Psi_P =\nabla_P \mathcal{E} +\frac1{2L} \Gamma_P\mathcal{E},
\end{equation}
where the $D$-dimensional spinor gauge parameter $\mathcal{E}(X)$ can again be decomposed as in \eqref{eq:RSgaugePar}. The Ricci tensor and scalar curvature for an Einstein manifold with a cosmological constant are \begin{equation}\label{eq:EinsteinDdim}
	\begin{split}
		R_{MN} &=\frac{2\Lambda}{D-2}G_{MN}, \\
		R &=\frac{2D}{D-2}\Lambda.
	\end{split}
\end{equation}
Consequently, the $d$- and $n$-dimensional manifolds must also be Einstein, with\footnote{This automatically excludes $n=1$, because there are no one-dimensional Einstein manifolds with negative curvature.}
\begin{equation}
	\begin{split}
		R^{(d)}_{\mu\nu} &=\frac{R^{(d)}}{d}g_{\mu\nu}, \quad\quad R^{(d)}=\frac{2d}{d+n-2}\Lambda, \\
		R^{(n)}_{pq} &=\frac{R^{(n)}}{n}\gamma_{pq}, \quad\quad R^{(n)}=\frac{2n}{d+n-2}\Lambda.
	\end{split}
\end{equation}

Let us derive the $d$-dimensional gauge transformations from \eqref{eq:spin32gauge10d}. Since $\mathcal{N}$ is Einstein, we can still use \eqref{eq:HodgeSpin32} as the KK decomposition for the spin-3/2 field. There is, however, an important difference: since $R^{(n)}_{pq}\ne0$, the $\Psi_i$ sum now includes harmonic $n$-dimensional spinors as well. That is because harmonic spinors are not parallel on a non-Ricci flat Einstein manifold.

The gauge transformations in $d$ dimensions are
\begin{equation}
	\begin{split}
		\delta\psi_{\mu,i} &=\nabla_\mu\epsilon_i +\frac1{2L}\gamma_\mu\epsilon_i, \\
		\delta\Phi_\Omega &=0, \\
		\delta\Psi_i &=\epsilon_i, \\
		\delta\Omega_i &=-\frac{in}{2L}\gamma_d\epsilon_i.
	\end{split}
\end{equation}
As before, the unitary gauge
\begin{equation}
	\Psi_\alpha =0, \quad\quad \Psi_i =0,
\end{equation}
is available, but now it includes the harmonic sector. Consequently, the gauge is completely fixed and there is no residual gauge symmetry in $d$ dimensions: $\psi_{\mu,\alpha}$ are actual massive spin-3/2 fields. This is reminiscent of the spontaneous supersymmetry breaking mechanism of 11D supergravity on the squashed seven-dimensional sphere, in which all gravitini can acquire a mass \cite{Duff:1983ajq}.

After gauge fixing, the action is the sum of \eqref{eq:SRpreShift} and
\begin{equation}
	S^{(m)}_{RS} =i\frac{D-2}{2L}\int d^dx\, e\, \Bigg[ -\sum_\Omega \bar\Phi_\Omega\Phi_\Omega +\sum_i \bigg(\bar\psi_{\mu,i}\gamma^{\mu\nu}\psi_{\nu,i}
	+i\big(\bar\psi_{\mu,i}\gamma^\mu\gamma_d\Omega_i -\bar\Omega_i\gamma_d\gamma^\mu\psi_{\mu,i}\big) +\frac{n-1}{n}\bar\Omega_i\Omega_i \bigg)\Bigg].
\end{equation}
The diagonalization of the action now requires a more convoluted shift of the fields, namely
\begin{equation}
	\psi_{\mu,i} \,\,\to\,\, \psi_{\mu,i} +\bigg(A\,\gamma_\mu +iB\,\gamma_\mu\gamma_d +C\,\nabla_\mu +iF\,\gamma_d\nabla_\mu \bigg)\Omega_i,
\end{equation}
valid for any value of $\lambda_i$, with
\begin{equation}
	\begin{split}
		A &=-\frac{D-2}{2nL(d-1)\Delta}\lambda_i, \quad\quad B=-\frac1{(d-1)\Delta}\bigg( \frac{n-1}{n}\lambda_i^2 +\frac{(n-1)(d-2)+n^2}{4L^2} \bigg), \\
		C &=-\frac{D-2}{n(d-1)\Delta}\lambda_i, \quad\quad F=\frac{D-2}{2L(d-1)\Delta}, \quad\quad \Delta =\lambda_i^2 +\frac{n\big( d(d-2)+n(d-1) \big)}{4(d-1)L^2}.
	\end{split}
\end{equation}
Note that $\Delta>0$. In the limit $L\to\infty$ these reduce to the shifts \eqref{eq:RSshift} of the massless case. With such shifts, the action becomes
\begin{equation}
	\begin{split}
		S_{mRS} &=\int d^dx\,e \bigg[ \sum_\Omega \Big(
		-\bar\Phi_\Omega i\gamma^\nu\nabla_\nu \Phi_\Omega
		-\lambda_\Omega\bar\Phi_\Omega\gamma_d\Phi_\Omega -i\frac{D-2}{2L} \bar\Phi_\Omega\Phi_\Omega \Big) \\
		&\quad\quad\quad\quad\quad\quad +\sum_i \bigg( -i\,\bar\psi_{\mu,i}\gamma^{\mu\nu\rho} \nabla_\nu \psi_{\rho,i} +\lambda_i\,\bar\psi_{\mu,i}\gamma_d\gamma^{\mu\nu}\psi_{\nu,i} +i\frac{D-2}{2L}\bar\psi_{\mu,i}\gamma^{\mu\nu}\psi_{\nu,i} \\
		&\quad\quad\quad\quad\quad\quad\quad  +N\Big(-\bar\Omega_i i\gamma^\mu \nabla_\mu \Omega_i +\lambda_i\bar\Omega_i\gamma_d\Omega_i -i\frac{D}{2L}\bar\Omega_i\Omega_i \Big) \bigg)\bigg],
	\end{split}
\end{equation}
where
\begin{equation}
	N =\frac{D-2}{(d-1)\Delta}\bigg( \frac{n-1}{n^2}\lambda_i^2 +\frac{D-1}{4L^2} \bigg) >0.
\end{equation}

To conclude, we perform the angle-dependent chirality rotations
\begin{equation}
	\Phi_\Omega \,\,\to\,\, e^{i\theta_\Omega\gamma_d}\Phi_\Omega, \quad\quad \psi_{\mu,i} \,\,\to\,\, e^{i\theta'_i\gamma_d}\psi_{\mu,i}, \quad\quad \Omega_i \,\,\to\,\, e^{i\theta_i\gamma_d}\Omega_i,
\end{equation}
with
\begin{equation}
	\begin{split}
		\sin(2\theta_\Omega) &=-\frac{\lambda_\Omega}{m_\Omega}, \quad\quad \cos(2\theta_\Omega) =-\frac{D-2}{2Lm_\Omega}, \quad\quad m_\Omega =\sqrt{\lambda^2_\Omega+\bigg(\frac{D-2}{2L}\bigg)^2}, \\
		\sin(2\theta'_i) &=\frac{\lambda_i}{m'_i}, \quad\quad \cos(2\theta'_i) =\frac{D-2}{2Lm'_i}, \quad\quad m'_i =\sqrt{\lambda^2_i+\bigg(\frac{D-2}{2L}\bigg)^2}, \\
		\sin(2\theta_i) &=\frac{\lambda_i}{m_i}, \quad\quad \cos(2\theta_i) =-\frac{D}{2Lm_i}, \quad\quad m_i =\sqrt{\lambda^2_i+\bigg(\frac{D}{2L}\bigg)^2},
	\end{split}
\end{equation}
and a trivial renormalization of $\Omega_i$. We arrive at the final form of the action:
\begin{equation}\label{eq:actionmRS}
	\begin{split}
		S_{mRS} &=-\int d^dx\,e \bigg[ \sum_\Omega \bar\Phi_\Omega \big(i\gamma^\mu\nabla_\mu -i\,m_\Omega \big)\Phi_\Omega \\
		&\quad\quad\quad\quad\quad\quad +\sum_i \bigg( \bar\psi_{\mu,i} \big( i\gamma^{\mu\nu\rho} \nabla_\nu -i\,m'_i\gamma^{\mu\rho} \big)\psi_{\rho,i} +\bar\Omega_i \big(i\gamma^\mu \nabla_\mu -i\,m_i \big)\Omega_i \bigg)\bigg].
	\end{split}
\end{equation}

The spectrum of the theory is now manifest, with
\begin{itemize}
	\item one massive spin-3/2 Dirac field $\psi_{\mu,i}$ for each eigenmode of the Dirac operator, with mass $m'_i=\sqrt{\lambda^2_i+\big((D-2)/2L\big)^2}$;
	\item one massive spin-1/2 Dirac field $\Omega_i$ for each eigenmode of the Dirac operator, with mass $m_i=\sqrt{\lambda^2_i+\big(D/2L\big)^2}$;
	\item one massive spin-1/2 Dirac field $\Phi_\Omega$ for each eigenmode of the RS operator, with mass $m_\Omega=\sqrt{\lambda^2_\Omega+\big((D-2)/2L\big)^2}$.
\end{itemize}
There are neither tachyons nor ghosts. In $d$ dimensions, all fields are massive, regardless of whether the corresponding eigenvalue in $n$ dimensions is zero. The action \eqref{eq:actionmRS} reduces to \eqref{eq:RSFinalKK} in the limit $L\to\infty$.

Let us conclude with a comment on $n=1,2$.
\paragraph{Special case: $\boldsymbol{n=2}$.}
For $n=2$, the sum over $\Omega$ is restricted to harmonic spinor-valued 1-forms $\Psi_{p,\omega}$. All formulas are regular at $n=2$. In this case, $\mathcal{N}$ is a compact surface of constant negative curvature --- i.e. a hyperbolic surface --- of genus $g\ge2$ with area $A=4\pi(g-1)L^2/(D-1)$ \cite{Besse}. The internal space can also be written as a quotient $\mathcal{N}=\mathbb{H}^2/\Gamma$, where $\Gamma$ is a torsion-free discrete subgroup of PSL$_2(\mathbb{R})$.

\paragraph{Special case: $\boldsymbol{n=1}$.}
As noted above, $n=1$ is incompatible with the condition that $\mathcal{N}$ is an Einstein manifold with negative curvature, and thus we can only consider $n>1$.

\section{Conclusions}\label{sec:concl}

In this paper we considered the Kaluza--Klein decomposition of spin-1/2 and spin-3/2 massless fields on a general product space and performed the dimensional reduction by keeping all massive modes. The result is a theory of infinite massive towers of particles in lower dimension, with potentially some massless fields as well. The analysis is done at the level of the action. The internal space is an arbitrary compact spin manifold, orientable and without boundary in the spin-1/2 case, while it is either Ricci-flat or Einstein in the spin-3/2 case, depending on whether the cosmological constant is zero or not.

By making use of a ``Hodge-like'' decomposition theorem for spinor-valued 1-forms, we were able to write down an ansatz for spin-3/2 fields on the internal space, which allowed us to work out the dimensional reduction explicitly. In the spin-3/2 case we also discussed gauge symmetry and kept track of its fate in the lower-dimensional theory. In particular, if a cosmological constant is present, the gauge symmetry is completely fixed and the lower-dimensional theory does not admit massless particles.

Let us now discuss some future directions. An interesting potential application of our results is the coupling of spin-1/2 and spin-3/2 fields to gravity. Given the KK decompositions, we can consider the minimal coupling of a spinor with gravity, expand both the $D$-dimensional metric $G_{MN}(X)$ (or vielbein $E_M^\mathscr{A}(X)$) and the spinor around a background, and expand at cubic or higher order in the fluctuations. At quadratic order, the spinor and the metric do not interact, because the fermionic background vanishes --- i.e. the spinor is ``pure fluctuation'' --- due to its Grassmann nature. The KK decomposition for the graviton $\delta G_{MN}=H_{MN}$ was worked out in \cite{Besse,Hinterbichler:2013kwa}, and thus one only needs to relate it to the vielbein fluctuation $\delta E_M^\mathscr{A}(X)$. This can be done e.g. by gauge fixing the local Lorentz transformations.

The KK decompositions for spin-1/2 and spin-3/2 fields discussed in this paper especially open the path to the study of dimensionally reduced supergravity theory and perturbative unitarity violations, in the spirit of \cite{Bonifacio:2019ioc,Bertucci:2026janew}. This would lead to sum rules involving eigenspectra of differential operators on some important 6D and 7D spin manifolds, like Calabi--Yau and $G_2$ manifolds.

Related to supergravity applications, throughout this paper we worked with Dirac fermions of spin 1/2 and 3/2. It is possible to perform similar KK decompositions for Weyl, Majorana and Majorana--Weyl spinors as well, for those dimensions which allow for such kinds of spinors.

It would be interesting to derive the KK decomposition for higher-spin fields, with spin $s$/2 for $s=5,7,\dots$, using the results of \cite{Homma:2020has}. Already for spin 5/2 --- i.e. symmetric rank-2 tensor-spinors ---, the constraints on the internal space $\mathcal{N}$ involve the Weyl tensor and are much stronger. This same obstruction shows up when trying to write a theory of interacting massless spin-5/2 fields coupled to gravity \cite{Fang:1978wz,Aragone:1979hx}. A possible way to circumvent these limitations is to consider Freund–Rubin-type flux compactifications \cite{Freund:1980xh} on e.g. AdS$_d\times S^n$, whose general KK ansatz was worked out in \cite{Hinterbichler:2013kwa}. A theory of interacting massive spin-5/2 fields on such background would serve as a toy model to study massive string theory excitations.

Another possible direction is the generalization of our results to warped product manifolds \cite{JCarot1993} --- i.e. spaces of the form $\mathcal{M}\times_f\mathcal{N}$ ---, whose interest in the context of dimensional reduction is well known \cite{Gogberashvili:1998vx,Randall:1999ee,Grossman:1999ra,Gherghetta:2000qt}. This would require constructing fermionic differential operators for arbitrary such spaces. Some progress was recently made for the scalar Laplacian in \cite{Hendrickson:2023wc}, following the construction of \cite{Witten:1982im}. The Dirac operator was analyzed in \cite{Law:2024pm}.

Finally, the bosonic KK ansatz of \cite{Hinterbichler:2013kwa} was recently used to compute the Euclidean gravity path integral at one loop for certain classes of product spaces \cite{Ivo:2025yek,Shi:2025amq}. Likewise, the spin-3/2 KK decomposition can be used to compute the Euclidean path integral in supergravity and higher-spin gravity theories defined on product spaces \cite{Turiaci:2025xwi,Anninos:2026hia,Baracco:2026ljd}.

\paragraph{Acknowledgements:}

We are grateful to James Bonifacio and Andrew Schrader for helpful discussions and collaboration on related topics, and especially to James Bonifacio for comments on the draft. We also thank Kurt Hinterbichler for useful discussions.

\appendix

\section{Clifford algebra in general dimensions}\label{app:clifford}

In this appendix we discuss flat space Clifford algebras. Dirac matrices in $D$ dimensions are capitalized, i.e. $\Gamma^\mathscr{A}$, while $d$-dimensional matrices are denoted by $\gamma^\mathfrak{a}$ and $n$-dimensional matrices are denoted by $\tilde\Gamma^{\bar{\mathfrak{a}}}$. The $2^{\lfloor D/2\rfloor}$-dimensional identity matrix in $D$ dimensions is denoted by $I_D$.

It is unfortunately not possible to pick a Clifford algebra representation which is valid for all values of $(d,n)$, with $D=d+n$. We thus focus on the case in which $d$ is even and $n$ is either even or odd.

The Clifford algebra in Minkowski spacetime in $D$ dimensions is
\begin{equation}
	\{\Gamma^\mathscr{A},\Gamma^\mathscr{B}\} = 2\eta^{\mathscr{AB}}I_D,
\end{equation}
for $\mathscr{A}=0,\dots,D-1$. When $D$ is even, one can also define the chirality matrix $\Gamma_D$ which satisfies
\begin{equation}
	\{\Gamma_D,\Gamma^\mathscr{A}\} = 0, \qquad \mathscr{A}=0,\dots,D-1.
\end{equation}
The Hermitian conjugate is
\begin{equation}
	\Gamma^{\mathscr{A}\dagger} =-\Gamma^0\Gamma^\mathscr{A} (\Gamma^0)^{-1}.
\end{equation}
Let us define the completely antisymmetrized product (with weight 1)
\begin{equation}
	\Gamma^{\mathscr{AB}\cdots \mathscr{C}} \coloneq \Gamma^{[\mathscr{A}} \Gamma^\mathscr{B} \cdots \Gamma^{\mathscr{C}]}.
\end{equation}
In $D$ dimensions, spinors have $2^{\lfloor D/2\rfloor}$ components. The Dirac adjoint of a spinor $\psi$ is
\begin{equation}
	\bar\psi \coloneq \psi^\dagger\Gamma^0.
\end{equation}

We can decompose $D$-dimensional indices as $\mathscr{A}=(\mathfrak{a},\bar{\mathfrak{a}})$, with $\mathfrak{a}=0,\dots,d-1$ and $\bar{\mathfrak{a}}=1,\dots,n$. We choose a representation of the matrices such that we can decompose the $D$-dimensional Clifford algebra as
\begin{equation}
	\begin{split}
		\Gamma^\mathfrak{a} &=\gamma^\mathfrak{a} \otimes I_n, \qquad \mathfrak{a}=0,\dots,d-1, \\
		\Gamma^{\bar{\mathfrak{a}}} &=\gamma_d \otimes \tilde\Gamma^{\bar{\mathfrak{a}}}, \qquad \bar{\mathfrak{a}}=1,\dots,n,
	\end{split}
\end{equation}
where $\gamma^\mathfrak{a}$ satisfy the Lorentzian Clifford algebra
\begin{equation}
	\{ \gamma^\mathfrak{a}, \gamma^\mathfrak{b}\} =2\eta^{\mathfrak{ab}}I_d,
\end{equation}
the chirality matrix $\gamma_d$ in $d$ dimensions satisfies
\begin{equation}
		\{ \gamma_d, \gamma^\mathfrak{a} \} =0, \qquad \mathfrak{a}=0,\dots,d-1,
\end{equation}
and $\tilde\Gamma^{\bar{\mathfrak{a}}}$ are Hermitian matrices which satisfy the Euclidean Clifford algebra
\begin{equation}
	\{\tilde\Gamma^{\bar{\mathfrak{a}}},\tilde\Gamma^{\bar{\mathfrak{b}}}\} =2\delta^{\bar{\mathfrak{a}}\bar{\mathfrak{b}}}I_n.
\end{equation}
If $n$ is even, we can also define a chirality matrix $\tilde\Gamma_n$ which satisfies
\begin{equation}
	\{ \tilde\Gamma_n,\tilde\Gamma^{\bar{\mathfrak{a}}} \} =0, \qquad \bar{\mathfrak{a}}=1,\dots,n.
\end{equation}

Let us conclude with some useful identities. Recall that $M,N,\dots$ denote (non-flat) spacetime indices in $D$-dimensions. On vectors $V^M$, spinors $\psi$ and spinor-valued 1-forms $\Psi_M$ the covariant derivatives are
\begin{equation}
	\begin{split}
		\nabla_M V^N &= \partial_M V^N +{\Gamma(G)^M}_{NP}V^P, \\
		\nabla_M \psi &= \partial_M \psi +\frac14 \Omega(E)_{M\mathscr{AB}}\Gamma^{\mathscr{AB}}\psi, \\
		\nabla_M \Psi_N &= \partial_M \Psi_N -{\Gamma(G)^P}_{MN}\Psi_P +\frac14 \Omega(E)_{M\mathscr{AB}}\Gamma^{\mathscr{AB}}\Psi_N,
	\end{split}
\end{equation}
where ${\Gamma(G)^M}_{NP}$ is the Christoffel connection and $\Omega(E)_{M\mathscr{AB}}$ is the spin connection. Thus we have
\begin{equation}
	\begin{split}
		[\nabla_M,\nabla_N] V^P &={R^P}_{QMN}V^Q, \\
		[\nabla_M,\nabla_N]\psi &=\frac14 R_{MNPQ}\Gamma^{PQ}\psi,
	\end{split}
\end{equation}
where ${R^P}_{QMN}$ is the Riemann tensor. We can also derive
\begin{equation}\label{eq:RiemannGammaId}
	R_{MNPQ}\Gamma^{MN}\Gamma^{PQ} =R_{MNPQ}\Gamma^M\Gamma^N\Gamma^P\Gamma^Q =-2R,
\end{equation}
where $R$ is the scalar curvature and we used the symmetries of the Riemann tensor and the Bianchi identity.

A standard computation shows that
\begin{equation}
	\big(\Gamma^M \nabla_M\big)^2 =\frac18 R_{MNPQ}\Gamma^{MN}\Gamma^{PQ} +\nabla_M\nabla^M
\end{equation}
on spinors. Using \eqref{eq:RiemannGammaId} we get the so-called Lichnerowicz--Weitzenb\"ock formula \cite{Lichnerowicz1963}
\begin{equation}\label{eq:LWformula}
	\big(\Gamma^M \nabla_M\big)^2 =-\frac R4 +\nabla_M\nabla^M.
\end{equation}
If a spinor satisfies the eigenvalue problem
\begin{equation}
	i\Gamma^M \nabla_M \psi =\lambda\psi,
\end{equation}
we conclude that
\begin{equation}
	-\nabla_M\nabla^M \psi =\left(\lambda^2-\frac R4\right)\psi.
\end{equation}

\section{Kaluza--Klein ansatz for spin 3/2}\label{app:SpinDecomp}

We want to analyze spin-3/2 --- or spinor-valued 1-forms --- fields on the $n$-dimensional ($n\ge2$) internal spin manifold $\mathcal{N}$, whose metric is denoted by $\gamma_{pq}$.

If $S_{1/2}$ denotes a spinor bundle on $\mathcal{N}$, spin-1/2 fields are sections of $S_{1/2}$, i.e. elements of $\Gamma(S_{1/2})$. Vectors are sections of a vector bundle $T\mathcal{N}$, or elements of $\Gamma(T\mathcal{N})$. Consider the famous decomposition
\begin{equation}\label{eq:gravitinoDecomp}
	S_{1/2} \otimes T\mathcal{N} = i(S_{1/2}) \oplus S_{3/2},
\end{equation}
where
\begin{equation}
	i:\Gamma(S_{1/2}) \to \Gamma(S_{1/2} \otimes T\mathcal{N}), \quad	\chi \mapsto \frac1n \tilde\Gamma_p \chi
\end{equation}
and $S_{3/2}$ represents ``gamma-traceless'' spin-3/2 fields, i.e. elements of the kernel of
\begin{equation}
	\mu:\Gamma(S_{1/2} \otimes T\mathcal{N}) \to \Gamma(S_{1/2}), \quad \mathcal{O}_p \mapsto \tilde\Gamma^p \mathcal{O}_p.
\end{equation}
For $\mathcal{O}_p \in \Gamma(S_{1/2} \otimes T\mathcal{N})$, the decomposition \eqref{eq:gravitinoDecomp} reads
\begin{equation}
	\mathcal{O}_p =\left( \mathcal{O}_p -\frac1n \tilde\Gamma_p \tilde\Gamma^q \mathcal{O}_q \right) +\frac1n \tilde\Gamma_p \tilde\Gamma^q \mathcal{O}_q.
\end{equation}
We also define the twistor operator and its adjoint,
\begin{equation}
	\begin{split}
		P:\Gamma(S_{1/2})\to \Gamma(S_{3/2}), &\quad \chi \mapsto \nabla_p \chi -\frac1n \tilde\Gamma_p \tilde\Gamma^q \nabla_q \chi, \\
		P^*:\Gamma(S_{3/2})\to \Gamma(S_{1/2}), &\quad \Psi_p \mapsto -\nabla_p \Psi^p.
	\end{split}
\end{equation}

Let $\mathcal{N}$ be compact. For $p$-forms we have the Hodge decomposition theorem; in this context we can write \cite{Wang_1991,Homma_2019}
\begin{equation}\label{eq:3/2dec}
	\Gamma(S_{3/2}) =\mathrm{Ker}P^* \oplus \mathrm{Im}P,
\end{equation}
where ${}^*$ denotes the adjoint. If $\mathcal{N}$ is Einstein, i.e. its Ricci tensor satisfies
\begin{equation}
	R^{(n)}_{pq} =\frac{R^{(n)}}n \gamma_{pq},
\end{equation}
the RS operator
\begin{equation}
	Q:\Gamma(S_{3/2}) \to \Gamma(S_{3/2}), \quad \Psi^p \mapsto i\tilde\Gamma^{pqr}\nabla_q\Psi_r -\frac{i}n\tilde\Gamma^p\tilde\Gamma_s \tilde\Gamma^{sqr}\nabla_q\Psi_r
\end{equation}
preserves the decomposition \eqref{eq:3/2dec} and we have
\begin{equation}\label{eq:QPDcompos}
	\begin{split}
		Q \circ P &=\frac{n-2}n P\circ \mathcal{D}, \\
		P^* \circ Q &=\frac{n-2}n \mathcal{D}\circ P^*.
	\end{split}
\end{equation}
We can therefore expand the elements of $\mathrm{Ker}P^* \oplus \mathrm{Im}P$ into eigenspaces of $Q$ and of the Dirac operator
\begin{equation}
	\mathcal{D}:\Gamma(S_{1/2}) \to \Gamma(S_{1/2}), \quad \chi\mapsto i\tilde\Gamma^p\nabla_p\chi.
\end{equation}
Note that $\mathcal{D}$ and $Q$ are both (formally) self-adjoint. Equivalently, we can consider the operator $Q_{phys}=i\tilde\Gamma^{pqr}\nabla_q$ and impose the transversality condition separately, as it is usually done in physics.

Let us assume $\mathcal{N}$ is Einstein. Then, if $\Psi_p \in \Gamma(S_{3/2})$ we write it as
\begin{equation}
	\Psi_p = \sum_\Omega c_\Omega \Psi_{p,\Omega} +\sum_i c_i \left( \nabla_p \chi_i -\frac1n \tilde\Gamma_p \tilde\Gamma^q \nabla_q \chi_i \right),
\end{equation}
where the spin-3/2 functions $\Psi_{p,\Omega} \in \mathrm{Ker}P^*$ form a basis satisfying
\begin{equation}\label{eq:PsipOmegaEqsApp}
	i\tilde\Gamma^q\nabla_q \Psi_{p,\Omega} =\lambda_\Omega \Psi_{p,\Omega}, \quad\quad	\tilde\Gamma^p \Psi_{p,\Omega} =\nabla^p \Psi_{p,\Omega} =0, \quad\quad \int_\mathcal{N} \Psi^\dagger_{p,\Omega}\Psi^p_{\Omega'} =\delta_{\Omega\Omega'}
\end{equation}
and $\chi_i \in \Gamma(S_{1/2})$ satisfy
\begin{equation}
	i\tilde\Gamma^{p}\nabla_p \chi_i =\lambda_i\chi_i, \quad\quad \int_\mathcal{N} \chi_i^\dagger\chi_j =\delta_{ij}.
\end{equation}

When $n=2$, $Q$ vanishes identically \cite{Homma_2019}, and $\mathrm{Ker}P^*$ is simply the finite-dimensional space given by spinor-valued 1-forms whose eigenvalue is zero, i.e. \eqref{eq:PsipOmegaEqsApp} with $\lambda_\Omega=0$. The case $n=1$ is pathological: in one dimension, $\tilde\Gamma^1=\pm1$, and both $Q$ and $P$ vanish identically; there are no ``gamma-traceless'' spinor-valued 1-forms.

\renewcommand{\em}{}
\bibliographystyle{utphys}
\addcontentsline{toc}{section}{References}
\bibliography{KK-fermions-refs}

\providecommand{\href}[2]{#2}\begingroup\raggedright\begin{thebibliography}{10}

\bibitem{Klein:1926tv}
O.~Klein, ``{Quantum Theory and Five-Dimensional Theory of Relativity},''
  \href{http://dx.doi.org/10.1007/BF01397481}{{\em Z. Phys.} {\bf 37} (1926)
  895--906}.

\bibitem{Kaluza:1921tu}
T.~Kaluza, ``{Zum Unit{\"a}tsproblem der Physik},''
  \href{http://dx.doi.org/10.1142/S0218271818700017}{{\em Sitzungsber. Preuss.
  Akad. Wiss. Berlin (Math. Phys.)} (1921)  966--972},
  \href{http://arxiv.org/abs/1803.08616}{{\tt arXiv:1803.08616
  [physics.hist-ph]}}.

\bibitem{Weyl:1918ib}
H.~Weyl, ``{Gravitation and electricity},'' {\em Sitzungsber. Preuss. Akad.
  Wiss. Berlin (Math. Phys.)} (1918)  465.

\bibitem{Salam:1981xd}
A.~Salam and J.~A. Strathdee, ``{On Kaluza-Klein Theory},''
  \href{http://dx.doi.org/10.1016/0003-4916(82)90291-3}{{\em Annals Phys.} {\bf
  141} (1982)  316--352}.

\bibitem{Witten:1981me}
E.~Witten, ``{Search for a Realistic Kaluza-Klein Theory*},''
  \href{http://dx.doi.org/10.1016/0550-3213(81)90021-3}{{\em Nucl. Phys. B}
  {\bf 186} (1981)  412}.

\bibitem{Bailin:1987jd}
D.~Bailin and A.~Love, ``{Kaluza-Klein Theories},''
  \href{http://dx.doi.org/10.1088/0034-4885/50/9/001}{{\em Rept. Prog. Phys.}
  {\bf 50} (1987)  1087--1170}.

\bibitem{Overduin:1997sri}
J.~M. Overduin and P.~S. Wesson, ``{Kaluza-Klein gravity},''
  \href{http://dx.doi.org/10.1016/S0370-1573(96)00046-4}{{\em Phys. Rept.} {\bf
  283} (1997)  303--380}, \href{http://arxiv.org/abs/gr-qc/9805018}{{\tt
  arXiv:gr-qc/9805018}}.

\bibitem{Nath:1975nj}
P.~Nath and R.~L. Arnowitt, ``{Generalized Supergauge Symmetry as a New
  Framework for Unified Gauge Theories},''
  \href{http://dx.doi.org/10.1016/0370-2693(75)90297-X}{{\em Phys. Lett. B}
  {\bf 56} (1975)  177--180}.

\bibitem{Freedman:1976xh}
D.~Z. Freedman, P.~van Nieuwenhuizen, and S.~Ferrara, ``{Progress Toward a
  Theory of Supergravity},''
  \href{http://dx.doi.org/10.1103/PhysRevD.13.3214}{{\em Phys. Rev. D} {\bf 13}
  (1976)  3214--3218}.

\bibitem{Freedman:1976py}
D.~Z. Freedman and P.~van Nieuwenhuizen, ``{Properties of Supergravity
  Theory},'' \href{http://dx.doi.org/10.1103/PhysRevD.14.912}{{\em Phys. Rev.
  D} {\bf 14} (1976)  912}.

\bibitem{Deser:1976eh}
S.~Deser and B.~Zumino, ``{Consistent Supergravity},''
  \href{http://dx.doi.org/10.1016/0370-2693(76)90089-7}{{\em Phys. Lett. B}
  {\bf 62} (1976)  335}.

\bibitem{VanNieuwenhuizen:1981ae}
P.~Van~Nieuwenhuizen, ``{Supergravity},''
  \href{http://dx.doi.org/10.1016/0370-1573(81)90157-5}{{\em Phys. Rept.} {\bf
  68} (1981)  189--398}.

\bibitem{Nahm:1977tg}
W.~Nahm, ``{Supersymmetries and Their Representations},''
  \href{http://dx.doi.org/10.1016/0550-3213(78)90218-3}{{\em Nucl. Phys. B}
  {\bf 135} (1978)  149}.

\bibitem{Cremmer:1978km}
E.~Cremmer, B.~Julia, and J.~Scherk, ``{Supergravity Theory in 11
  Dimensions},'' \href{http://dx.doi.org/10.1016/0370-2693(78)90894-8}{{\em
  Phys. Lett. B} {\bf 76} (1978)  409--412}.

\bibitem{Duff:1986hr}
M.~J. Duff, B.~E.~W. Nilsson, and C.~N. Pope, ``{Kaluza-Klein Supergravity},''
  \href{http://dx.doi.org/10.1016/0370-1573(86)90163-8}{{\em Phys. Rept.} {\bf
  130} (1986)  1--142}.

\bibitem{Duff:2025tot}
M.~J. Duff, B.~E.~W. Nilsson, and C.~N. Pope, ``{Kaluza-Klein Supergravity
  2025},'' \href{http://arxiv.org/abs/2502.07710}{{\tt arXiv:2502.07710
  [hep-th]}}.

\bibitem{Rarita:1941mf}
W.~Rarita and J.~Schwinger, ``{On a theory of particles with half integral
  spin},'' \href{http://dx.doi.org/10.1103/PhysRev.60.61}{{\em Phys. Rev.} {\bf
  60} (1941)  61}.

\bibitem{Velo:1969bt}
G.~Velo and D.~Zwanziger, ``{Propagation and quantization of Rarita-Schwinger
  waves in an external electromagnetic potential},''
  \href{http://dx.doi.org/10.1103/PhysRev.186.1337}{{\em Phys. Rev.} {\bf 186}
  (1969)  1337--1341}.

\bibitem{Alvarez-Gaume:1983ihn}
L.~Alvarez-Gaume and E.~Witten, ``{Gravitational Anomalies},''
  \href{http://dx.doi.org/10.1016/0550-3213(84)90066-X}{{\em Nucl. Phys. B}
  {\bf 234} (1984)  269}.

\bibitem{Grisaru:1976vm}
M.~T. Grisaru, H.~N. Pendleton, and P.~van Nieuwenhuizen, ``{Supergravity and
  the S Matrix},'' \href{http://dx.doi.org/10.1103/PhysRevD.15.996}{{\em Phys.
  Rev. D} {\bf 15} (1977)  996}.

\bibitem{Casher:1984ym}
A.~Casher, F.~Englert, H.~Nicolai, and M.~Rooman, ``{The Mass Spectrum of
  Supergravity on the Round Seven Sphere},''
  \href{http://dx.doi.org/10.1016/0550-3213(84)90392-4}{{\em Nucl. Phys. B}
  {\bf 243} (1984)  173}.

\bibitem{Fogleman:1984jd}
G.~Fogleman and K.~S. Viswanathan, ``{Harmonic Expansions for (4+K)-Dimensional
  Rarita-Schwinger Fields on Coset Spaces and Effective Lagrangians in
  Four-Dimensions},'' \href{http://dx.doi.org/10.1103/PhysRevD.31.299}{{\em
  Phys. Rev. D} {\bf 31} (1985)  299--306}.

\bibitem{Kim:1985ez}
H.~J. Kim, L.~J. Romans, and P.~van Nieuwenhuizen, ``{The Mass Spectrum of
  Chiral N=2 D=10 Supergravity on $S^5$},''
  \href{http://dx.doi.org/10.1103/PhysRevD.32.389}{{\em Phys. Rev. D} {\bf 32}
  (1985)  389}.

\bibitem{Hinterbichler:2013kwa}
K.~Hinterbichler, J.~Levin, and C.~Zukowski, ``{Kaluza-Klein Towers on General
  Manifolds},'' \href{http://dx.doi.org/10.1103/PhysRevD.89.086007}{{\em Phys.
  Rev. D} {\bf 89} (2014) no.~8, 086007},
  \href{http://arxiv.org/abs/1310.6353}{{\tt arXiv:1310.6353 [hep-th]}}.

\bibitem{Bonifacio:2019ioc}
J.~Bonifacio and K.~Hinterbichler, ``{Unitarization from Geometry},''
  \href{http://dx.doi.org/10.1007/JHEP12(2019)165}{{\em JHEP} {\bf 12} (2019)
  165}, \href{http://arxiv.org/abs/1910.04767}{{\tt arXiv:1910.04767
  [hep-th]}}.

\bibitem{Bertucci:2026janew}
{F. Bertucci, J. Bonifacio, and A. Schrader,} {\textit{to appear.}}

\bibitem{Belavin:1984vu}
A.~A. Belavin, A.~M. Polyakov, and A.~B. Zamolodchikov, ``{Infinite Conformal
  Symmetry in Two-Dimensional Quantum Field Theory},''
  \href{http://dx.doi.org/10.1016/0550-3213(84)90052-X}{{\em Nucl. Phys. B}
  {\bf 241} (1984)  333--380}.

\bibitem{Rattazzi:2008pe}
R.~Rattazzi, V.~S. Rychkov, E.~Tonni, and A.~Vichi, ``{Bounding scalar operator
  dimensions in 4D CFT},''
  \href{http://dx.doi.org/10.1088/1126-6708/2008/12/031}{{\em JHEP} {\bf 12}
  (2008)  031},
\href{http://arxiv.org/abs/0807.0004}{{\tt arXiv:0807.0004 [hep-th]}}.

\bibitem{Poland:2018epd}
D.~Poland, S.~Rychkov, and A.~Vichi, ``{The Conformal Bootstrap: Theory,
  Numerical Techniques, and Applications},''
  \href{http://dx.doi.org/10.1103/RevModPhys.91.015002}{{\em Rev. Mod. Phys.}
  {\bf 91} (2019)  015002}, \href{http://arxiv.org/abs/1805.04405}{{\tt
  arXiv:1805.04405 [hep-th]}}.

\bibitem{Simmons-Duffin:2015qma}
D.~Simmons-Duffin, ``{A Semidefinite Program Solver for the Conformal
  Bootstrap},'' \href{http://dx.doi.org/10.1007/JHEP06(2015)174}{{\em JHEP}
  {\bf 06} (2015)  174}, \href{http://arxiv.org/abs/1502.02033}{{\tt
  arXiv:1502.02033 [hep-th]}}.

\bibitem{Bonifacio:2020xoc}
J.~Bonifacio and K.~Hinterbichler, ``{Bootstrap Bounds on Closed Einstein
  Manifolds},'' \href{http://dx.doi.org/10.1007/JHEP10(2020)069}{{\em JHEP}
  {\bf 10} (2020)  069}, \href{http://arxiv.org/abs/2007.10337}{{\tt
  arXiv:2007.10337 [hep-th]}}.

\bibitem{Bonifacio:2021msa}
J.~Bonifacio, ``{Bootstrap bounds on closed hyperbolic manifolds},''
  \href{http://dx.doi.org/10.1007/JHEP02(2022)025}{{\em JHEP} {\bf 02} (2022)
  025}, \href{http://arxiv.org/abs/2107.09674}{{\tt arXiv:2107.09674
  [hep-th]}}.

\bibitem{Bonifacio:2021aqf}
J.~Bonifacio, ``{Bootstrapping closed hyperbolic surfaces},''
  \href{http://dx.doi.org/10.1007/JHEP03(2022)093}{{\em JHEP} {\bf 03} (2022)
  093}, \href{http://arxiv.org/abs/2111.13215}{{\tt arXiv:2111.13215
  [hep-th]}}.

\bibitem{Radcliffe:2024jcg}
A.~Radcliffe, ``{Non-saturation of bootstrap bounds by hyperbolic orbifolds},''
  \href{http://dx.doi.org/10.1007/JHEP12(2025)115}{{\em JHEP} {\bf 12} (2025)
  115}, \href{http://arxiv.org/abs/2404.14479}{{\tt arXiv:2404.14479
  [hep-th]}}.

\bibitem{Adve:2025rvf}
A.~Adve, J.~Bonifacio, P.~Kravchuk, D.~Mazac, S.~Pal, A.~Radcliffe, and
  G.~Rogelberg, ``{Weyl bound for trilinear periods via conformal bootstrap},''
  \href{http://arxiv.org/abs/2508.20576}{{\tt arXiv:2508.20576 [math.NT]}}.

\bibitem{Kravchuk:2021akc}
P.~Kravchuk, D.~Mazac, and S.~Pal, ``{Automorphic spectra and the conformal
  bootstrap},'' \href{http://dx.doi.org/10.1090/cams/26}{{\em Commun. Am. Math.
  Soc.} {\bf 4} (2024) no.~1, 1--63},
  \href{http://arxiv.org/abs/2111.12716}{{\tt arXiv:2111.12716 [hep-th]}}.

\bibitem{Bonifacio:2023ban}
J.~Bonifacio, D.~Mazac, and S.~Pal, ``{Spectral Bounds on Hyperbolic
  3-Manifolds: Associativity and the Trace Formula},''
  \href{http://dx.doi.org/10.1007/s00220-024-05222-0}{{\em Commun. Math. Phys.}
  {\bf 406} (2025)  51}, \href{http://arxiv.org/abs/2308.11174}{{\tt
  arXiv:2308.11174 [math.SP]}}.

\bibitem{Bertucci:2026fjh}
F.~Bertucci and J.~Bonifacio, ``{Bootstrapping Euclidean Lattices},''
  \href{http://arxiv.org/abs/2606.13771}{{\tt arXiv:2606.13771 [hep-th]}}.

\bibitem{Gesteau:2023brw}
E.~Gesteau, S.~Pal, D.~Simmons-Duffin, and Y.~Xu, ``{Bounds on spectral gaps of
  Hyperbolic spin surfaces},''
  \href{http://dx.doi.org/10.56994/JAMR.003.001.003}{{\em J. Assoc. Math. Res.}
  {\bf 3} (2025) no.~1, 72--139}, \href{http://arxiv.org/abs/2311.13330}{{\tt
  arXiv:2311.13330 [math.SP]}}.

\bibitem{Hitchin:1974rbi}
N.~Hitchin, ``{Harmonic Spinors},''
  \href{http://dx.doi.org/10.1016/0001-8708(74)90021-8}{{\em Adv. Math.} {\bf
  14} (1974) no.~1, 1--55}.

\bibitem{baer2000}
C.~B\"ar, ``{Dependence on the spin structure of the Dirac spectrum},''
  \href{http://arxiv.org/abs/math/0007131}{{\tt arXiv:math/0007131 [math.DG]}}.

\bibitem{Mecklenburg:1981ju}
W.~Mecklenburg, ``{On the Zero Modes of the Internal Dirac Operator in
  {Kaluza-Klein} Theories},''
  \href{http://dx.doi.org/10.1103/PhysRevD.26.1327}{{\em Phys. Rev. D} {\bf 26}
  (1982)  1327}.

\bibitem{Candelas:1985en}
P.~Candelas, G.~T. Horowitz, A.~Strominger, and E.~Witten, ``{Vacuum
  configurations for superstrings},''
  \href{http://dx.doi.org/10.1016/0550-3213(85)90602-9}{{\em Nucl. Phys. B}
  {\bf 258} (1985)  46--74}.

\bibitem{Wang_1991}
M.~Y. Wang, ``Preserving parallel spinors under metric deformations,'' {\em
  Indiana University Mathematics Journal} {\bf 40} (1991) no.~3, 815--844.

\bibitem{Homma_2019}
Y.~Homma and U.~Semmelmann, ``{The Kernel of the Rarita–Schwinger Operator on
  Riemannian Spin Manifolds},''
  \href{http://dx.doi.org/10.1007/s00220-019-03324-8}{{\em Communications in
  Mathematical Physics} {\bf 370} (2019) no.~3, 853–871}.

\bibitem{Deser:1977uq}
S.~Deser and B.~Zumino, ``{Broken Supersymmetry and Supergravity},''
  \href{http://dx.doi.org/10.1103/PhysRevLett.38.1433}{{\em Phys. Rev. Lett.}
  {\bf 38} (1977)  1433--1436}.

\bibitem{Weinberg:2000cr}
S.~Weinberg, {\em {The quantum theory of fields. Vol. 3: Supersymmetry}}.
\newblock Cambridge University Press, 2013.

\bibitem{Buchbinder:2010gp}
I.~L. Buchbinder and V.~A. Krykhtin, ``{BRST Lagrangian construction for
  spin-3/2 field in Einstein space},''
  \href{http://dx.doi.org/10.1142/S0217732310033372}{{\em Mod. Phys. Lett. A}
  {\bf 25} (2010)  1667--1677}, \href{http://arxiv.org/abs/1003.0185}{{\tt
  arXiv:1003.0185 [hep-th]}}.

\bibitem{Rahman:2011ik}
R.~Rahman, ``{Helicity-1/2 mode as a probe of interactions of a massive
  Rarita-Schwinger field},''
  \href{http://dx.doi.org/10.1103/PhysRevD.87.065030}{{\em Phys. Rev. D} {\bf
  87} (2013) no.~6, 065030}, \href{http://arxiv.org/abs/1111.3366}{{\tt
  arXiv:1111.3366 [hep-th]}}.

\bibitem{Deser:2001pe}
S.~Deser and A.~Waldron, ``{Gauge invariances and phases of massive higher
  spins in (A)dS},''
  \href{http://dx.doi.org/10.1103/PhysRevLett.87.031601}{{\em Phys. Rev. Lett.}
  {\bf 87} (2001)  031601}, \href{http://arxiv.org/abs/hep-th/0102166}{{\tt
  arXiv:hep-th/0102166}}.

\bibitem{Deser:2001us}
S.~Deser and A.~Waldron, ``{Partial masslessness of higher spins in (A)dS},''
  \href{http://dx.doi.org/10.1016/S0550-3213(01)00212-7}{{\em Nucl. Phys. B}
  {\bf 607} (2001)  577--604}, \href{http://arxiv.org/abs/hep-th/0103198}{{\tt
  arXiv:hep-th/0103198}}.

\bibitem{Bergshoeff:2015tra}
E.~A. Bergshoeff, D.~Z. Freedman, R.~Kallosh, and A.~Van~Proeyen, ``{Pure de
  Sitter Supergravity},''
  \href{http://dx.doi.org/10.1103/PhysRevD.93.069901}{{\em Phys. Rev. D} {\bf
  92} (2015) no.~8, 085040}, \href{http://arxiv.org/abs/1507.08264}{{\tt
  arXiv:1507.08264 [hep-th]}}. [Erratum: Phys.Rev.D 93, 069901 (2016)].

\bibitem{Duff:1983ajq}
M.~J. Duff, B.~E.~W. Nilsson, and C.~N. Pope, ``{Spontaneous Supersymmetry
  Breaking by the Squashed Seven Sphere},''
  \href{http://dx.doi.org/10.1103/PhysRevLett.50.2043}{{\em Phys. Rev. Lett.}
  {\bf 50} (1983) no.~26, 2043--2046}. [Erratum: Phys.Rev.Lett. 51, 846
  (1983)].

\bibitem{Besse}
A.~L. Besse, {\em {Einstein Manifolds}}.
\newblock Springer-Verlag, Berlin, Heidelberg, New York,
1987.
\newblock

\bibitem{Homma:2020has}
Y.~Homma and T.~Tomihisa, ``{The spinor and tensor fields with higher spin on
  spaces of constant curvature},''
  \href{http://dx.doi.org/10.1007/s10455-021-09791-4}{{\em Annals Global Anal.
  Geom.} {\bf 60} (2021) no.~4, 829--861},
  \href{http://arxiv.org/abs/2005.09840}{{\tt arXiv:2005.09840 [math.DG]}}.

\bibitem{Fang:1978wz}
J.~Fang and C.~Fronsdal, ``{Massless Fields with Half Integral Spin},''
  \href{http://dx.doi.org/10.1103/PhysRevD.18.3630}{{\em Phys. Rev. D} {\bf 18}
  (1978)  3630}.

\bibitem{Aragone:1979hx}
C.~Aragone and S.~Deser, ``{Consistency Problems of Hypergravity},''
  \href{http://dx.doi.org/10.1016/0370-2693(79)90808-6}{{\em Phys. Lett. B}
  {\bf 86} (1979)  161--163}.

\bibitem{Freund:1980xh}
P.~G.~O. Freund and M.~A. Rubin, ``{Dynamics of Dimensional Reduction},''
  \href{http://dx.doi.org/10.1016/0370-2693(80)90590-0}{{\em Phys. Lett. B}
  {\bf 97} (1980)  233--235}.

\bibitem{JCarot1993}
J.~Carot and J.~da~Costa, ``{On the geometry of warped spacetimes},''
  \href{http://dx.doi.org/10.1088/0264-9381/10/3/007}{{\em Classical and
  Quantum Gravity} {\bf 10} (1993) no.~3, 461}.

\bibitem{Gogberashvili:1998vx}
M.~Gogberashvili, ``{Hierarchy problem in the shell universe model},''
  \href{http://dx.doi.org/10.1142/S0218271802002992}{{\em Int. J. Mod. Phys. D}
  {\bf 11} (2002)  1635--1638}, \href{http://arxiv.org/abs/hep-ph/9812296}{{\tt
  arXiv:hep-ph/9812296}}.

\bibitem{Randall:1999ee}
L.~Randall and R.~Sundrum, ``{A Large mass hierarchy from a small extra
  dimension},'' \href{http://dx.doi.org/10.1103/PhysRevLett.83.3370}{{\em Phys.
  Rev. Lett.} {\bf 83} (1999)  3370--3373},
  \href{http://arxiv.org/abs/hep-ph/9905221}{{\tt arXiv:hep-ph/9905221}}.

\bibitem{Grossman:1999ra}
Y.~Grossman and M.~Neubert, ``{Neutrino masses and mixings in nonfactorizable
  geometry},'' \href{http://dx.doi.org/10.1016/S0370-2693(00)00054-X}{{\em
  Phys. Lett. B} {\bf 474} (2000)  361--371},
  \href{http://arxiv.org/abs/hep-ph/9912408}{{\tt arXiv:hep-ph/9912408}}.

\bibitem{Gherghetta:2000qt}
T.~Gherghetta and A.~Pomarol, ``{Bulk fields and supersymmetry in a slice of
  AdS},'' \href{http://dx.doi.org/10.1016/S0550-3213(00)00392-8}{{\em Nucl.
  Phys. B} {\bf 586} (2000)  141--162},
  \href{http://arxiv.org/abs/hep-ph/0003129}{{\tt arXiv:hep-ph/0003129}}.

\bibitem{Hendrickson:2023wc}
K.~Hendrickson, ``{Warped Compactification on General Manifolds}.'' {Master
  Thesis}, 2023.
\newblock \url{https://hdl.handle.net/11299/259542}.

\bibitem{Witten:1982im}
E.~Witten, ``{Supersymmetry and Morse theory},'' {\em J. Diff. Geom.} {\bf 17}
  (1982) no.~4, 661--692.

\bibitem{Law:2024pm}
M.~B. Law, I.~M. Lopez, and D.~Santiago, ``{Positive mass and Dirac operators
  on weighted manifolds and smooth metric measure spaces},''
  \href{http://arxiv.org/abs/2312.15441}{{\tt arXiv:2312.15441 [math.DG]}}.

\bibitem{Ivo:2025yek}
V.~Ivo, J.~Maldacena, and Z.~Sun, ``{Physical instabilities and the phase of
  the Euclidean path integral},''
  \href{http://dx.doi.org/10.1007/JHEP04(2026)118}{{\em JHEP} {\bf 04} (2026)
  118}, \href{http://arxiv.org/abs/2504.00920}{{\tt arXiv:2504.00920
  [hep-th]}}.

\bibitem{Shi:2025amq}
X.~Shi and G.~J. Turiaci, ``{The phase of the gravitational path integral},''
  \href{http://dx.doi.org/10.1007/JHEP07(2025)047}{{\em JHEP} {\bf 07} (2025)
  047}, \href{http://arxiv.org/abs/2504.00900}{{\tt arXiv:2504.00900
  [hep-th]}}.

\bibitem{Turiaci:2025xwi}
G.~J. Turiaci and C.-H. Wu, ``{The wavefunction of a quantum S$^{1}$
  {\texttimes} S$^{2}$ universe},''
  \href{http://dx.doi.org/10.1007/JHEP07(2025)158}{{\em JHEP} {\bf 07} (2025)
  158}, \href{http://arxiv.org/abs/2503.14639}{{\tt arXiv:2503.14639
  [hep-th]}}.

\bibitem{Anninos:2026hia}
D.~Anninos, C.~Baracco, V.~A. Letsios, and B.~M{\"u}hlmann, ``{dS$^{4}$
  Metamorphosis},'' \href{http://dx.doi.org/10.1007/JHEP07(2026)178}{{\em JHEP}
  {\bf 07} (2026)  178}, \href{http://arxiv.org/abs/2602.19812}{{\tt
  arXiv:2602.19812 [hep-th]}}.

\bibitem{Baracco:2026ljd}
C.~Baracco and V.~A. Letsios, ``{Sphere path integrals for fermionic gauge
  fields},'' \href{http://arxiv.org/abs/2609.19275}{{\tt arXiv:2609.19275
  [hep-th]}}.

\bibitem{Lichnerowicz1963}
A.~Lichnerowicz, ``{Spineurs harmoniques},'' {\em C. R. Acad. Sci. Paris} {\bf
  257} (1963)  7--9.

\end{thebibliography}\endgroup

\end{document}